\documentclass[twocolumn,showpacs,prl]{revtex4}

\usepackage{graphicx}
\usepackage{amssymb}
\usepackage{epstopdf}
\usepackage{epsfig}

\begin{document}

\def\aprge{\buildrel > \over {_{\sim}}}
\def\aprle{\buildrel < \over {_{\sim}}}

\def\etal{{\it et.~al.}}
\def\ie{{\it i.e.}}
\def\eg{{\it e.g.}}

\def\bwt{\begin{widetext}}
\def\ewt{\end{widetext}}
\def\be{\begin{equation}}
\def\ee{\end{equation}}
\def\bea{\begin{eqnarray}}
\def\eea{\end{eqnarray}}
\def\bean{\begin{eqnarray*}}
\def\eean{\end{eqnarray*}}
\def\bary{\begin{array}}
\def\eary{\end{array}}
\def\bi{\bibitem}
\def\bit{\begin{itemize}}
\def\eit{\end{itemize}}

\def\lan{\langle}
\def\ran{\rangle}
\def\lra{\leftrightarrow}
\def\la{\leftarrow}
\def\ra{\rightarrow}
\def\dash{\mbox{-}}
\def\ol{\overline}

\def\ub{\ol{u}}
\def\db{\ol{d}}
\def\sb{\ol{s}}
\def\cb{\ol{c}}

\def\re{\rm Re}
\def\im{\rm Im}

\def \b{{\cal B}}
\def \ca{{\cal A}}
\def \ko{K^0}
\def \ok{\overline{K}^0}
\def \s{\sqrt{2}}
\def \st{\sqrt{3}}
\def \sx{\sqrt{6}}

\hspace{14.0cm}\hfill {\tt CAS-KITPC/ITP-086}\\

\title{{\bf The Lepton-Number-Violating  Decays of $B^+, D^+ $ and $D_s^+$ Mesons Induced by the Doubly Charged Higgs Boson}}

\author{Yong-Liang Ma}


\address{Institute of High Energy Physics, CAS, P.O.Box 918(4), Beijing 100049, China\\
Theoretical Physics Center for Science Facilities, CAS, Beijing 100049\\
Institute of Theoretical Physics, Chinese Academy of sciences, Beijing 100190\\
Kavli Institute for Theoretical Physics China, Chinese Academy of
Science, Beijing 100190, China}
\date{\today}
\begin{abstract}
The lepton-number-violaing decays of $B^+, D^+$ and $D_s^+$ mesons
induced by the doubly charged Higgs boson have been studied. It is
found that although the yielded results of the branch ratio are much
smaller than the present limits from the data they are consistent
with the previous conclusions calculated in the framework of the
relativistic quark model where the processes happened via light
Majorana neutrinos.
\end{abstract}
 \maketitle

%


Exploring the physics beyond the standard model is a hot topic in
particle physics. To this aim, some extensions of the standard model
were made and a typical kind of these models are the left-right
symmetric
models~\cite{Pati:1974yy,Mohapatra:1974hk,Mohapatra:1974gc,Senjanovic:1975rk,Marshak:1979fm}.
Recently, the left-right symmetric models were extended to
incorporate the dark matter candidate~\cite{Guo:2008hy,Guo:2008si}.
One of the common characters of these models is that a doubly
charged Higgs boson was introduced in a triplet Higgs
representation. Because of the existence of this doubly charged
Higgs boson many new phenomenologies arise, for example the
lepton-number-violating and lepton-flavor-violating processes that
will be studied in this paper.

The phenomenologies relevant to the doubly charged Higgs boson have
been investigated both theoretically and experimentally. For example
in Refs.~\cite{Georgi:1985nv,Vega:1989tt} the production of the
doubly charged Higgs boson $H^{++}$ was sudied. In
Ref.~\cite{Picciotto:1997tk}, the role of the doubly charged Higgs
boson in the lepton-number-violating decay of $K^+ \to \pi^\pm
\mu^\mp \mu^\mp$ was studied and it was found that although the
decay rate due to the doubly charged Higgs boson is very small it is
of the same order of magnitude as the rate for kaon double $\beta$
decay induced by light or heavy Majorana neutrinos. Experimentally,
the physics of the doubly charged Higgs boson have been performed by
many collaborations as mentioned in the following references.
Considering this, we will study the lepton-number-violating decays
of bottom and charm mesons in this paper. This study is meaningful
because some experiments such as the BESIII will improve the
measurements of the branch ratio of some lepton-number-violating
decay processes to the order of $10^{-9}$.

The general form of the lepton-number-violating coupling to
left-handed leptons is specified by the following Lagrangian:
\begin{eqnarray}
{\cal L}_{\rm int} & = & i h_{ij}\psi^T_{iL}C\sigma_2\Delta
\psi_{jL} + {\rm H.c.}, \label{LdoubleH}
\end{eqnarray}
where $h_{ij}(i,j = 1,2,3)$ are arbitrary coupling constants,
$\sigma_2$ is the Pauli matrix, $C$ is the Dirac charge conjugation
operator, $\psi_{iL}$ is the ith generation left-handed lepton
doublet, and $\Delta$ is the $2 \times 2$ representation of the $Y =
2$ complex triplet. Explicitly,
\begin{eqnarray}
\psi_{iL} & = & \left(
               \begin{array}{c}
                 v_i \\
                 l_i \\
               \end{array}
             \right)_{L} \; ; ~~~~
\Delta = \left(
               \begin{array}{cc}
                 H^-/\sqrt{2} & H^{--} \\
                 H^0 & -H^-/\sqrt{2} \\
               \end{array}
             \right).
\end{eqnarray}
It should be noted that in the left-right symmetric models $\Delta$
should be specified as $\Delta_L$ and the left-handed gauge symmetry
was specified as the standard model ${\rm SU}(2)_L$ gauge symmetry.

From the Lagrangian (\ref{LdoubleH}), one can get the decay rate for
$H^{\pm\pm} \to l^{\pm}_i l^{\pm}_j$ as
\begin{eqnarray}
\Gamma(H^{\pm\pm} \to l^{\pm}_i l^{\pm}_j) & = &
C_{ij}\frac{h_{ij}^2}{8\pi s\sqrt{s}}
\lambda^{1/2}(s,m_i^2,m_j^2)\nonumber\\
& & \times (s - m_i^2 - m_j^2), \label{widthll}
\end{eqnarray}
where $C_{ij} = 1(2)$ for $i = j(i \neq j)$ and $s$ is the invariant
momentum square transferred to the final leptons $l_i$ and $l_j$. In
the case of real doubly charged Higgs boson $H^{++}$, $s =
m_{H^{++}}^2$. When the final leptons are both massless, the width
(\ref{widthll}) can be simplified as
\begin{eqnarray}
\Gamma(H^{\pm\pm} \to l^{\pm}_i l^{\pm}_j) & = &
C_{ij}\frac{h_{ij}^2}{8\pi } m_{H^{++}}.
\end{eqnarray}

For the coupling constants $h_{ij}$, the present experiments give
the following constraints. From the Bhabha scattering one obtains
the limit for $h_{ee}$ as~\cite{Swartz:1989qz}
\begin{eqnarray}
h_{ee}^2 & \simeq & 9.7 \times 10^{-6} {\rm GeV}^{-2} M_{H^{--}}^2.
\label{hee}
\end{eqnarray}
The $(g-2)_\mu$ measurement~\cite{Brown:2001mga} provides an upper
limit for $h_{\mu\mu}$ as
\begin{eqnarray}
h_{\mu\mu}^2 & \simeq & 2.5 \times 10^{-5} {\rm GeV}^{-2}
M_{H^{--}}^2. \label{hmm}
\end{eqnarray}
For the flavor changing interaction, the most stringent constraint
comes from the upper limit for the flavor changing decay $\mu \to
\bar{e} e e $ which puts the following limit on the relevant
coupling constants:
\begin{eqnarray}
h_{e\mu}h_{ee} & \leq & 3.2 \times 10^{-11} {\rm GeV}^{-2}
M_{H^{--}}^2\label{hemhee}
\end{eqnarray}
and the nonobservation of the decay $\mu \to e \gamma $ leads to
\begin{eqnarray}
h_{e\mu}h_{\mu\mu} & \leq & 2.0 \times 10^{-10} {\rm GeV}^{-2}
M_{H^{--}}^2. \label{hemhmm}
\end{eqnarray}
From the Bhabba scattering with LEP data, the flavor violating
coupling constants are found as~\cite{Atag:2004dg}
\begin{eqnarray}
h_{e(\mu,\tau)}^2 & \leq & 1.0 \times 10^{-6} {\rm GeV}^{-2}
M_{H^{--}}^2.
\end{eqnarray}
In the following calculation we will adopt the upper limits of the
relevant coupling constants. Combining (\ref{hmm}) with
(\ref{hemhmm}) one can get
\begin{eqnarray}
h_{e\mu}^2 & \leq & 1.6 \times 10^{-15} {\rm GeV}^{-2} M_{H^{--}}^2.
\label{hem}
\end{eqnarray}
This numerical value means, compared with the
lepton-flavor-concerving processes, the lepton-flavor-violating
processes are dramatically suppressed.

The mass of doubly charged Higgs boson $H^{++}$ has been searched by
several
collaborations~\cite{Abbiendi:2003pr,Achard:2003mv,Abdallah:2002qj,Acosta:2004uj,Abazov:2004au}.
The data have excluded $H^{++}$ boson below the mass of about
$100~$GeV by assuming exclusive $H^{++}$ decays to a given dilepton
channel. And the search performed by the CDF and D0 Collaborations
at the Fermi Tevatron in the $\mu\mu$ channel have excluded the
$H^{++}$ below a mass of $136(113)~$GeV~\cite{Acosta:2004uj} and
$118~$GeV~\cite{Abazov:2004au}. In the following calculation, we
will adopt the lower limit of $H^{++}$ mass, that is, adopt
$m_{H^{++}} = 100~$GeV.

\begin{figure}[htbp]
\begin{center}
\includegraphics[scale=0.4]{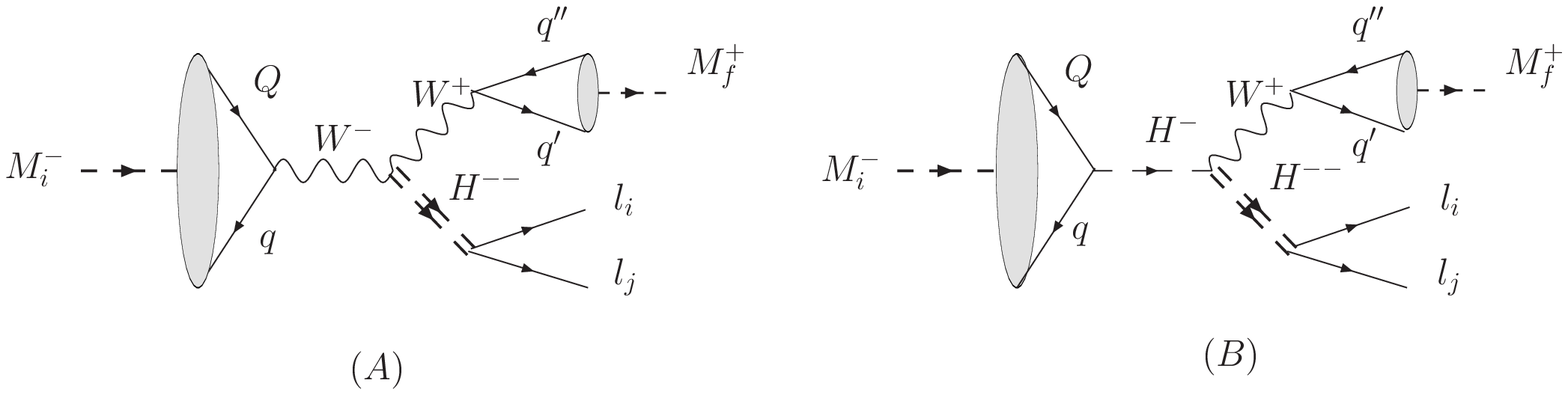}\\
\includegraphics[scale=0.4]{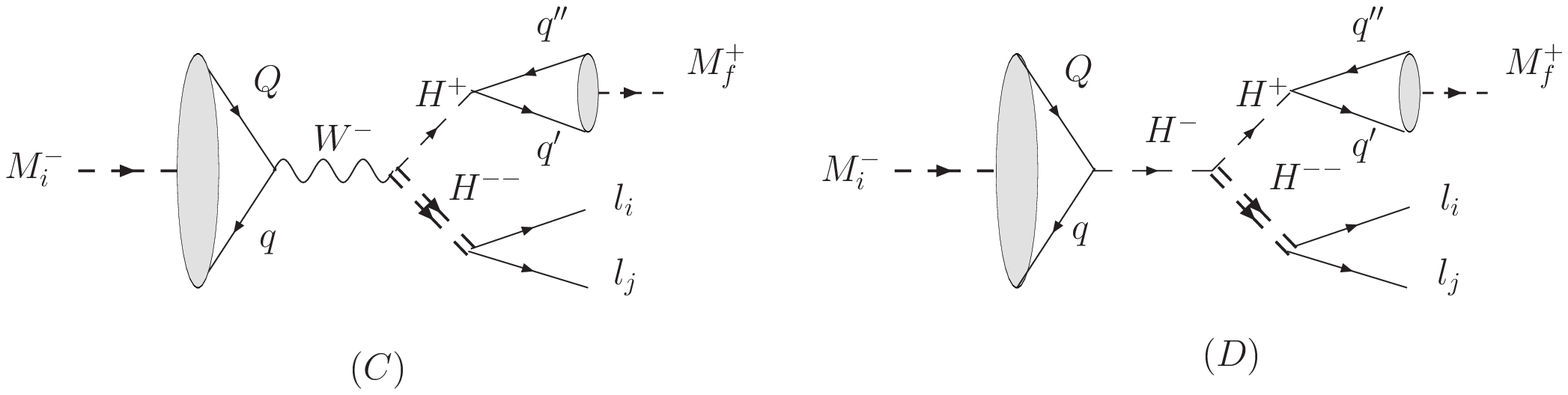}\\
\includegraphics[scale=0.4]{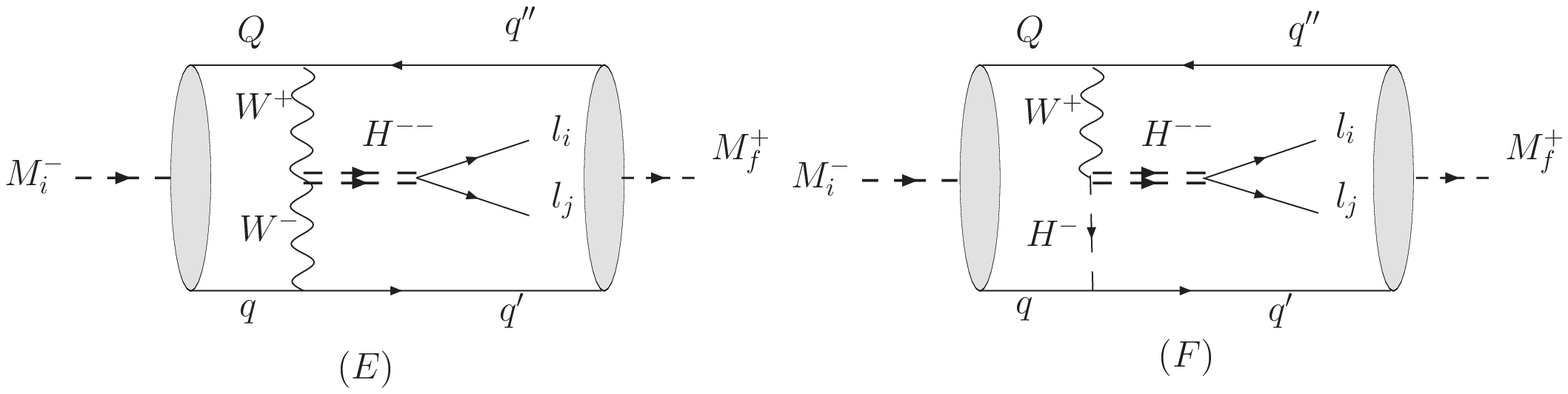}\\
\includegraphics[scale=0.4]{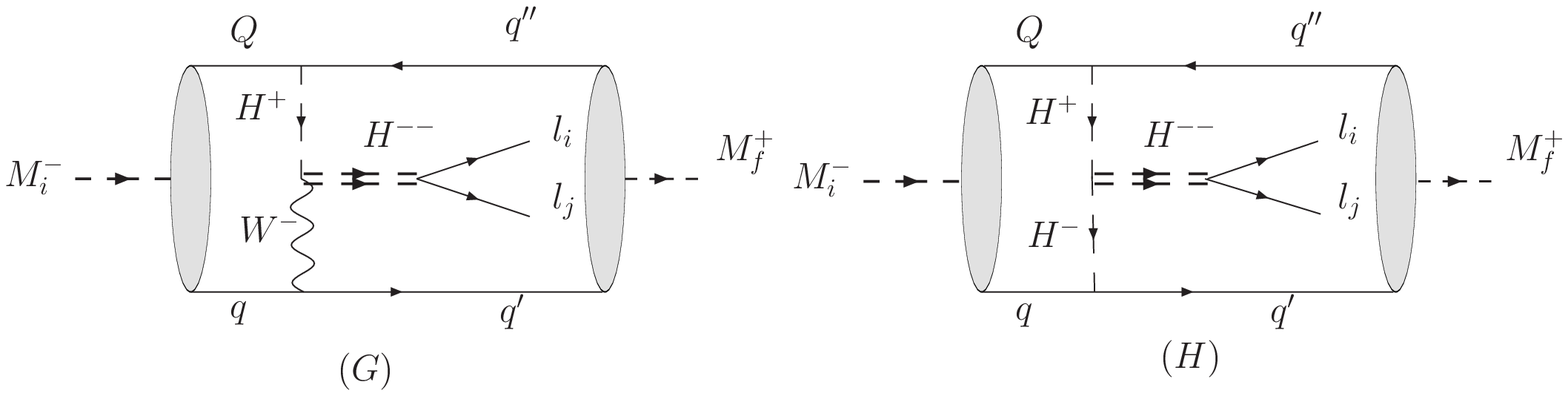}
\end{center}
\caption[Diagrams contribute to the decay of $M_i^- \to M_f^+ l_i l_j $. $M_i$ is the initial heavy mesons and $M_f^+$ is the pseudoscalar or vector mesons in the final states.]{%
Diagrams contribute to the decay of $M_i^- \to M_f^+ l_i l_j $.
$M_i$ is the initial heavy mesons and $M_f^+$ is the pseudoscalar or
vector meson in the final states. } \label{fig:decay}
\end{figure}

Next, we will calculate the lepton-number-violating decays of some
heavy mesons induced by the doubly charged Higgs boson $H^{++}$,
i.e., the decays of $M_{(s)}^- \to P^+(V^+) l_i l_j$. To calculate
these decay processes, the diagrams at quark level shown in
Fig.\ref{fig:decay} should be considered.

For the relevant leptonic decay constants, we will adopt the
following ordinary definitions:
\begin{eqnarray}
\langle 0| \bar{q}\gamma_\mu \gamma_5 q^\prime |P(p) \rangle & = &-i p_\mu f_{P} \nonumber\\
\langle 0| \bar{q}\gamma_\mu q^\prime |V(p) \rangle & = &
\epsilon_\mu m_V f_{V}, \label{leptondecayc}
\end{eqnarray}
where $P$ and $V$ denote pseudoscalar and vector mesons,
respectively. $\epsilon_\mu$ is the polarization vector of relevant
vector meson and $m_V$ is the vector meson mass. The numerical
values of the leptonic decay constants can be yielded from the
relevant processes as
~\cite{Wise:1992hn,Amsler:2008zz,Beneke:2003zv,Ball:1998ff,Buchalla:2008jp,Gray:2005ad}
\begin{eqnarray}
f_\pi & = & 135 {~\rm MeV}; ~~~~~~ f_K = 160 {~\rm MeV}; \nonumber\\
f_{\rho^+} & = & 209 {~\rm MeV}; ~~~~ f_{K^{\ast+}} = 218 {~\rm MeV};\nonumber\\
f_{D^+} & = & 222.6 {~\rm MeV}; ~~~~ f_{D_s^+} = 280.1 {~\rm MeV};
\nonumber\\
f_{B^+} & = & 216 {~\rm MeV}. \nonumber
\end{eqnarray}

To calculate the decay widths, the following interaction Lagrangian
besides the Lagrangian (\ref{LdoubleH}) should be applied:
\begin{eqnarray}
{\cal L}_{\rm int} & = & {\cal L}_{\rm int}^{\rm SM} + {\cal L}_{\rm
int}^{\rm BSM} \nonumber\\
{\cal L}_{\rm int}^{\rm SM} & = & \frac{g}{\sqrt{2}} V_{ud} W_\mu^+ \bar{u}_L\gamma_\mu d_L + h.c. \nonumber\\
{\cal L}_{\rm int}^{\rm BSM} & = & - \sqrt{2} g m_W s_H g_{\mu\nu} H^{++} W_\mu^- W_\mu ^- \nonumber\\
& & + \frac{\sqrt{2}}{2} c_H g W_\mu^- H^- \partial^{^{^{\hspace{-0.2cm}\leftrightarrow}}}_\mu H^{++} \nonumber\\
& & + \frac{ig s_H}{\sqrt{2}m_W c_H} H^+(m_{q^\prime}\bar{q}_R
q^\prime_R - m_q \bar{q}_L q^\prime_L ) \nonumber\\
& & + 3 \sqrt{2}s_H \bar{\lambda}_{45}v H^{++} H^{-} H^{-} + h.c.
\label{LintBSM}
\end{eqnarray}
where $g = e/\sin\theta_W$ with $\theta_W$ as the Weinberg angle,
$V_{ud}$ is the Cabibbo-Kobayashi-Maskawa matrix element and $c_H$
and $s_H$ are the cosine and sine of the mixing angle and it was
found the current data lead to the limit $s_H \leq 0.0056$ at $95\%$
confidence level~\cite{Chakrabarti:1998qy}. To write down ${\cal
L}_{\rm int}^{\rm BSM}$ we have applied the convention of
Ref.~\cite{Gunion:1989ci}.

As an example, we will consider the lepton-number-violating decay of
$B^- \to \pi^+ l_i l_j $. In this case, $Q = b, q = \bar{u},
q^\prime = u$, and $q^{\prime\prime} = \bar{d}$ in
Fig.~\ref{fig:decay}. After analyzing the analytic forms of the
matrix elements, one can see the contributions from diagrams $(A)$
and $(E)$ are dominant. The argument is the following: According to
the Lagrangian (\ref{LintBSM}), the matrix element of the first
diagram is proportional to $m_W\sin_H$, while that of the second
diagram is proportional to $c_H\frac{s_H m_q~p}{m_WC_W} = \frac{s_H
m_q~p}{m_W}$, where $p$ would be maximally a few GeV and $m_{q}$ is
the quark mass, so the second diagram is suppressed. A similar
analysis can be made on diagram ~(C). For diagram $(D)$, it is even
suppressed relative to diagram ~(B) concerning the Lagrangian
(\ref{LintBSM}). So that, for the first four diagrams, contribution
from the diagram~(A) is dominant. Along the same reasoning, for the
last four diagrams, the contribution from the diagram~(E) is
dominant. Then we have the dominant contributions as
\begin{eqnarray}
iM_{(A)} & = & \frac{\sqrt{2}g^3}{2}V_{ub}^{\ast}V_{ud}
\nonumber\\
& & \times \langle \pi^+(p_\pi)| [\bar{b}\gamma_\mu P_L
u][\bar{u}\gamma_\mu P_L d] |
B^-(p_B)\rangle \nonumber\\
& & \times \frac{1}{m_{W}^3 m_{H^{++}}^2} s_H
h_{l_il_j}\langle {\rm leptons} \rangle,  \\
iM_{(E)} & = & \Big (\frac{1}{3} \Big
)\frac{\sqrt{2}g^3}{2}V_{ub}^{\ast}V_{ud}
\nonumber\\
& & \times \langle \pi^+(p_\pi)| [\bar{d}\gamma_\mu P_L u]
[\bar{b}\gamma_\mu P_L u] |
B^-(p_B)\rangle \nonumber\\
& & \times \frac{1}{m_{W}^3 m_{H^{++}}^2} s_H h_{l_il_j}\langle {\rm
leptons} \rangle,
\end{eqnarray}
where the factor $(1/3)$ is from the Fierz transformations of Dirac
indices and color indices and $\langle {\rm leptons} \rangle =
\bar{v}^s(p_{i})P_Lu^s(p_j) - (p_i \leftrightarrow p_j)$.

Using the leptonic constants defined in (\ref{leptondecayc}), we can
rewrite the matric elements as
\begin{eqnarray}
iM & = & iM_{(A)} + iM_{(E)} \label{sumele}\\
& = & \frac{\sqrt{2}g^3 s_H}{6m_{W}^3
m_{H^{++}}^2}V_{ub}^{\ast}V_{ud}f_B f_\pi p_B\cdot p_\pi
h_{l_il_j}\langle {\rm leptons} \rangle, \nonumber
\end{eqnarray}
which leads to
\begin{eqnarray}
|iM|^2 & = & \frac{g^6|V_{ub}|^2|V_{ud}|^2 f_\pi^2 f_B^2
s_H^2}{18m_{W}^6
m_{H^{++}}^4} \nonumber\\
& & \times (p_B \cdot p_\pi)^2 h_{l_il_j}^2\sum_s|\langle {\rm
leptons} \rangle|^2,
\end{eqnarray}
where $s$ is the spins of the final leptons.

With this expression one can write the decay width of $B^- \to \pi^+
l_i l_j $ as
\begin{eqnarray}
& & \frac{d\Gamma(B^- \to \pi^+ l_i l_j )}{ds} \nonumber\\
& & \;\;\;\;\;\;\; =
\frac{1}{2m_B}\int\frac{d^3p_\pi}{(2\pi)^3}\frac{1}{2E_\pi}\int\frac{d^3p_{l_i}}{(2\pi)^3}\frac{1}{2E_{l_i}}\int\frac{d^3p_{l_j}}{(2\pi)^3}\frac{1}{2E_{l_j}}\nonumber\\
& & \;\;\;\;\;\;\;\;\;\;\;\; \times |M|^2 (2\pi)^4\delta^4(q -
p_{l_i} -
p_{l_j})\delta[q^2-(p_B - p_\pi)^2] \nonumber \\
& & \;\;\;\;\;\;\; = C_{ij}\Big(\frac{h_{l_i
l_j}s_H}{m_{H^{++}}^2}\Big)^2\frac{\alpha_{\rm em}^3
|V_{ub}|^2|V_{ud}|^2f_\pi^2 f_B^2}{144\sin^6\theta_W
m_B^3m_{W}^6}   \nonumber\\
& & \;\;\;\;\;\;\;\;\;\;\;\; \times
\lambda^{1/2}(s,m_i^2,m_j^2)\lambda^{1/2}(s,m_B^2,m_\pi^2)
\nonumber\\
& & \;\;\;\;\;\;\;\;\;\;\;\;  \times\frac{(s - m_B^2 - m_\pi^2)^2(s
- m_i^2 - m_j^2)}{s}, \label{GammaBpi}
\end{eqnarray}
where $\lambda$ is the K$\ddot{a}$llen function, $m_i$ is the mass
of ith flavor lepton. $ s $ is the invariant momentum square
transferred to the leptons and its amplitude is between the region
$(m_{l_i} + m_{l_j})^2 \leq s \leq (m_B - m_\pi)^2$. Along the same
method, one can easily write down the decay width of $B^- \to \rho^+
l_i l_j $ as
\begin{eqnarray}
& & \frac{d\Gamma(B^- \to \rho^+ l_i l_j)}{ds} \nonumber\\
& & \;\;\;\;\;\;\;\;\; = C_{ij}\Big(\frac{h_{l_i
l_j}s_H}{m_{H^{++}}^2}\Big)^2\frac{\alpha_{\rm em}^3
|V_{ub}|^2|V_{ud}|^2f_\rho^2 f_B^2}{144\sin^6\theta_W
m_B^3m_{W}^6}   \nonumber\\
& & \;\;\;\;\;\;\;\;\;\;\;\; \times
\lambda^{1/2}(s,m_i^2,m_j^2)\lambda^{3/2}(s,m_B^2,m_\rho^2)
\nonumber\\
& & \;\;\;\;\;\;\;\;\;\;\;\;  \times\frac{(s - m_i^2 - m_j^2)}{s},
\label{GammaBrho}
\end{eqnarray}
where the region of $ s $ is $(m_{l_i} + m_{l_j})^2 \leq s \leq (m_B
- m_\rho)^2$ in this case.

Substituting the relevant physical quantities, one can get the
following numerical results:
\begin{eqnarray}
\Gamma(B^- \to \pi^+ e^- e^- ) & < & 5.80 \times 10^{-15}{~\rm eV} \nonumber\\
\Gamma(B^- \to \rho^+ e^- e^- ) & < & 1.04 \times 10^{-14}{~\rm eV}.
\end{eqnarray}
In Tables~\ref{table:resultsb}, \ref{table:resultsd}, and
\ref{table:resultsds}, we list our numerical results for the branch
ratio of all the channels that we are interested in and the
corresponding data from PDG~\cite{Amsler:2008zz}. From the tables
one can see that the branch ratio induced by the doubly charged
Higgs boson is very small. Our results also show that the
lepton-flavor-violating decays are dramatically suppressed which
agree with our above expectation. Compared with the pseudoscalar
channel, the corresponding vector meson channel is improved because
of the larger leptonic decay constant of the relevant vector meson.
And among all the three initial states, $B^+, D^+$, and $D_s^+$, the
branch ratio of $D_s^+$ channel is the most important. The reason is
that the Cabibbo-Kobayashi-Maskawa matrix element $V_{cs}$ is the
largest one among $V_{ub}, V_{cd}$, and $V_{cs}$.

At last, we would like to say that, in the second identity of
(\ref{sumele}) the naive factorization has been applied. When the
QCD corrections are included, minor corrections should be made to
our numerical results. For example, for the charm sector, the
improvement of the branch ratio is about 20\%, but for the bottom
sector the corrections are much smaller and can even be neglected.
Concerning the present status of the relevant experiments, our
results are meaningful and when the precision of the experiments are
improved one should calculate the QCD corrections explicitly.

In summary, in this paper we studied the lepton-number-violating
decays of $B^+, D^+$ and $D_s^+$ mesons induced by the doubly
charged Higgs boson $H^{++}$. It is found that the branch ratio we
yielded here are much smaller than the present limits from
corresponding data. It should be noted that our conclusion is
consistent with that of Ref.~\cite{Picciotto:1997tk}, where the
lepton-number-violaing decay of $K^\pm \to \pi^\mp \mu^\pm \mu^\pm$
induced by doubly charged Higgs boson was studied.

\begin{table}

\caption{\label{table:resultsb} The branch ratio of the
lepton-number-violating B meson decays.}

\begin{tabular}{lll}
\hline\hline \hspace*{.3cm} Decay modes \hspace*{.3cm}
& \hspace*{.2cm} Present results \hspace*{.2cm}& \hspace*{.2cm} Data~\cite{Amsler:2008zz} \hspace*{.2cm} \\
\hline
\,\,\, $B^- \to \pi^+ e^-e^- $ & \,\,\,\,\, $ < 5.81 \times 10^{-24} $ \,\,\,\,\,& \,\,\,\,\, $ < 1.6 \times 10^{-6}$  \,\,\, \\
\,\,\, $B^- \to \pi^+ \mu^-\mu^- $ & \,\,\,\,\, $ < 1.49 \times 10^{-23} $ \,\,\,\,\,& \,\,\,\,\, $ < 1.4 \times 10^{-6}$ \,\,\, \\
\,\,\, $B^- \to \pi^+ e^-\mu^- $ & \,\,\,\,\, $ < 1.91 \times 10^{-33} $ \,\,\,\,\,& \,\,\,\,\, $ < 1.3 \times 10^{-6}$ \,\,\, \\
\,\,\, $B^- \to K^+ e^-e^- $ & \,\,\,\,\, $ < 4.31 \times 10^{-25} $ \,\,\,\,\,& \,\,\,\,\, $ < 1.0 \times 10^{-6}$  \,\,\, \\
\,\,\, $B^- \to K^+ \mu^-\mu^- $ & \,\,\,\,\, $ < 1.10 \times 10^{-24} $ \,\,\,\,\,& \,\,\,\,\, $ < 1.8 \times 10^{-6}$ \,\,\, \\
\,\,\, $B^- \to K^+ e^-\mu^- $ & \,\,\,\,\, $ < 1.41 \times 10^{-34} $ \,\,\,\,\,& \,\,\,\,\, $ < 2.0 \times 10^{-6}$ \,\,\, \\
\,\,\, $B^- \to \rho^+ e^-e^- $ & \,\,\,\,\, $ < 1.04 \times 10^{-23} $ \,\,\,\,\,& \,\,\,\,\, $ < 2.6 \times 10^{-6}$ \,\,\, \\
\,\,\, $B^- \to \rho^+ \mu^-\mu^- $ & \,\,\,\,\, $ < 2.67 \times 10^{-23} $ \,\,\,\,\,& \,\,\,\,\, $ < 5.0 \times 10^{-6}$ \,\,\, \\
\,\,\, $B^- \to \rho^+ e^-\mu^- $ & \,\,\,\,\, $ < 3.43 \times 10^{-33} $ \,\,\,\,\,& \,\,\,\,\, $ < 3.3 \times 10^{-6}$ \,\,\, \\
\,\,\, $B^- \to K^{\ast+} e^-e^- $ & \,\,\,\,\, $ < 5.60 \times 10^{-25} $ \,\,\,\,\,& \,\,\,\,\, $ < 2.8 \times 10^{-6}$  \,\,\, \\
\,\,\, $B^- \to K^{\ast+} \mu^-\mu^- $ & \,\,\,\,\, $ < 1.43 \times 10^{-24} $ \,\,\,\,\,& \,\,\,\,\, $ < 8.3 \times 10^{-6}$  \,\,\, \\
\,\,\, $B^- \to K^{\ast+} e^-\mu^- $ & \,\,\,\,\, $ < 1.84 \times 10^{-34} $ \,\,\,\,\,& \,\,\,\,\, $ < 4.4 \times 10^{-6}$ \,\,\, \\
\hline\hline
\end{tabular}
\end{table}

\begin{table}

\caption{\label{table:resultsd}  The branch ratio of the
lepton-number-violating D meson decays. }

\begin{tabular}{lll}
\hline \hline \hspace*{.4cm} Decay modes \hspace*{.4cm}
& \hspace*{.2cm} Present results \hspace*{.1cm}& \hspace*{.2cm} Data~\cite{Amsler:2008zz} \hspace*{.2cm} \\
\hline
\,\,\, $D^- \to \pi^+ e^-e^- $ & \,\,\,\,\, $ < 7.71 \times 10^{-24} $ \,\,\,\,\,& \,\,\,\, $ < $ 3.6 $\times 10^{-6}$  \,\,\, \\
\,\,\, $D^- \to \pi^+ \mu^-\mu^- $ & \,\,\,\,\, $ < 1.86 \times 10^{-23} $ \,\,\,\,\,& \,\,\,\, $ < $ 4.8 $\times 10^{-6}$  \,\,\, \\
\,\,\, $D^- \to \pi^+ e^-\mu^- $ & \,\,\,\,\, $ < 2.46 \times 10^{-33} $ \,\,\,\,\,& \,\,\,\, $ < $ 5.0 $\times 10^{-5}$  \,\,\, \\
\,\,\, $D^- \to K^+ e^-e^- $ & \,\,\,\,\, $ < 4.63 \times 10^{-25} $ \,\,\,\,\,& \,\,\,\, $ < $ 4.5 $\times 10^{-6}$  \,\,\, \\
\,\,\, $D^- \to K^+ \mu^-\mu^- $ & \,\,\,\,\, $ < 1.11 \times 10^{-24} $ \,\,\,\,\,& \,\,\,\, $ < $ 1.3 $\times 10^{-5}$  \,\,\, \\
\,\,\, $D^- \to K^+ e^-\mu^- $ & \,\,\,\,\, $ < 1.47 \times 10^{-34} $ \,\,\,\,\,& \,\,\,\, $ < $ 1.3 $\times 10^{-4}$  \,\,\, \\
\,\,\, $D^- \to \rho^+ e^-e^- $ & \,\,\,\,\, $ < 2.52 \times 10^{-24} $ \,\,\,\,\,& \,\,\,\, ~  \,\,\,\,\, \\
\,\,\, $D^- \to \rho^+ \mu^-\mu^- $ & \,\,\,\,\, $ < 5.68 \times 10^{-24} $ \,\,\,\,\,& \,\,\,\, $ < $ 5.6 $\times 10^{-4}$  \,\,\, \\
\,\,\, $D^- \to \rho^+ e^-\mu^- $ & \,\,\,\,\, $ < 7.79 \times 10^{-34} $ \,\,\,\,\,& \,\,\,\, ~  \,\,\, \\
\,\,\, $D^- \to K^{\ast+} e^-e^- $ & \,\,\,\,\, $ < 7.90 \times 10^{-26} $ \,\,\,\,\,& \,\,\,\, ~  \,\,\, \\
\,\,\, $D^- \to K^{\ast+} \mu^-\mu^- $ & \,\,\,\,\, $ < 1.72 \times 10^{-25} $ \,\,\,\,\,& \,\,\,\, $ < $ 8.5 $\times 10^{-4}$  \,\,\, \\
\,\,\, $D^- \to K^{\ast+} e^-\mu^- $ & \,\,\,\,\, $ < 2.40 \times 10^{-35} $ \,\,\,\,\,& \,\,\,\, ~  \,\,\, \\
\hline\hline
\end{tabular}
\end{table}

\begin{table}

\caption{\label{table:resultsds}  The branch ratio of the
lepton-number-violating $D_s$ meson decays. }

\begin{tabular}{lll}
\hline \hline \hspace*{.4cm} Decay modes \hspace*{.4cm}
& \hspace*{.2cm} Present results \hspace*{.1cm}& \hspace*{.2cm} Data~\cite{Amsler:2008zz} \hspace*{.2cm} \\
\hline
\,\,\, $D_s^- \to \pi^+ e^-e^- $ & \,\,\,\,\, $ < 1.46 \times 10^{-22} $ \,\,\,\,\,& \,\,\,\, $ < $ 6.9 $\times 10^{-4}$  \,\,\, \\
\,\,\, $D_s^- \to \pi^+ \mu^-\mu^- $ & \,\,\,\,\, $ < 3.55 \times 10^{-22} $ \,\,\,\,\,& \,\,\,\, $ < $ 2.9 $\times 10^{-5}$  \,\,\, \\
\,\,\, $D_s^- \to \pi^+ e^-\mu^- $ & \,\,\,\,\, $ < 4.68 \times 10^{-32} $ \,\,\,\,\,& \,\,\,\, $ < $ 7.3 $\times 10^{-4}$  \,\,\, \\
\,\,\, $D_s^- \to K^+ e^-e^- $ & \,\,\,\,\, $ < 9.02 \times 10^{-24} $ \,\,\,\,\,& \,\,\,\, $ < $ 6.3 $\times 10^{-4}$  \,\,\, \\
\,\,\, $D_s^- \to K^+ \mu^-\mu^- $ & \,\,\,\,\, $ < 2.17 \times 10^{-23} $ \,\,\,\,\,& \,\,\,\, $ < $ 1.3 $\times 10^{-5}$  \,\,\, \\
\,\,\, $D_s^- \to K^+ e^-\mu^- $ & \,\,\,\,\, $ < 2.88 \times 10^{-33} $ \,\,\,\,\,& \,\,\,\, $ < $ 6.8 $\times 10^{-4}$  \,\,\, \\
\,\,\, $D_s^- \to \rho^+ e^-e^- $ & \,\,\,\,\,\, $ < 5.76 \times 10^{-23} $ \,\,\,\,\,& \,\,\,\, ~  \,\,\, \\
\,\,\, $D_s^- \to \rho^+ \mu^-\mu^- $ & \,\,\,\,\, $ < 1.32 \times 10^{-22} $ \,\,\,\,\,& \,\,\,\, ~  \,\,\, \\
\,\,\, $D_s^- \to \rho^+ e^-\mu^- $ & \,\,\,\,\, $ < 1.80 \times 10^{-32} $ \,\,\,\,\,& \,\,\,\, ~  \,\,\, \\
\,\,\, $D_s^- \to K^{\ast+} e^-e^- $ & \,\,\,\,\, $ < 1.92 \times 10^{-24} $  \,\,\,\,\,& \,\,\,\, ~  \,\,\, \\
\,\,\, $D_s^- \to K^{\ast+} \mu^-\mu^- $ & \,\,\,\,\, $ < 4.32 \times 10^{-24} $  \,\,\,\,\,& \,\,\,\, $ < $ 1.4 $\times 10^{-3}$  \,\,\, \\
\,\,\, $D_s^- \to K^{\ast+} e^-\mu^- $ & \,\,\,\,\, $ < 5.92 \times 10^{-34} $ \,\,\,\,\,& \,\,\,\, ~  \,\,\, \\
\hline \hline
\end{tabular}
\end{table}




\acknowledgments


\vspace{0.5cm}

We would like to thank W-L Guo for his valuable discussions on the
left-right symmetry models. This work was supported in part by the
Project of Knowledge Innovation Program (PKIP) of Chinese Academy of
Sciences under Contract No.KJCX2-YW-W10.



\end{document}